\documentstyle[a4,12pt,epsfig,graphicx]{article}
\begin{document}
\titlepage
\begin{flushright}
CERN-TH/2001-006\\
T/01-010\\
hep-th/0101077 \\
\end{flushright}
\vskip 1cm
\begin{center}
{ \Large
\bf Graviton Absorption by Non-BPS Branes}
\end{center}

\vskip 1cm
\begin{center}
{\large Ph. Brax\footnote{email: philippe.brax@cern.ch} }
\end{center}
\vskip 0.5cm
\begin{center}
Theoretical Physics Division, CERN\\
CH-1211 Geneva 23\footnote { On leave of absence from  Service de Physique Th\'eorique, 
CEA-Saclay F-91191 Gif/Yvette Cedex, France}\\
\end{center}

\vskip 2cm
\begin{center}
{\bf  Abstract}
\end{center}
\vskip .2 cm
\noindent

\vskip .3in 
\baselineskip10pt{
We consider the behaviour of neutral non-BPS branes probed by scalars
and gravitons. We show that the naked singularity of the non-BPS branes
is a {\it repulson} absorbing no incoming radiation. The naked singularity
is surrounded by an infinite potential well breaking the unitarity of the
scattering S-matrix. We compute the absorption cross section which
is  infrared divergent. In particular this
confirms that gravity does not decouple for  the non-BPS branes.  
}
\bigskip
\vskip 1 cm

\noindent
\newpage
\baselineskip=1.5\baselineskip

\section{Introduction}
Non-BPS branes have been extensively studied recently both from the
theoretical and phenomenological points of view\cite{senreview}.
In string theories the non-BPS branes appear to be unstable
configuration in dimensions at odds with the usual BPS branes, i.e.
$2p$-branes in type IIB strings and $(2p+1)$-branes in type IIA strings.
One of the salient features of this construction based upon orbifolding 
coincident brane-antibrane pairs is the presence of a real tachyonic field
with a non-trivial potential\cite{sentach}. In particular the fate of the non-BPS branes
has been linked to the rolling of the tachyon along its decreasing potential
and its eventual condensation at the minimum of the potential.
The precise description of the potential has been tackled using 
string field theory methods\cite{numerics}.
Recent  advances in describing the decay of the unstable branes 
have appeared  from the point of view of matrix theory and 
non-commutative geometry\cite{Harvey-Horava-Kraus,das,har,man}.
Extensions to unstable M-branes have also been discussed\cite{intri}.

Another approach to the issue of non-BPS branes comes from the supergravity side
where exact solutions have been obtained and analysed\cite{ler,brax}. They describe
charged and neutral configurations. The parameter space of Poincar\'e invariant
solutions comprises
three directions readily identified with the ADM mass, the charge and an
extra integration constant $c_1$ which appears to be a function of the tachyonic vev. The decay of the unstable branes can be seen as a path in the parameter space\cite{brax}.

An intriguing feature of these solutions is the presence of a naked singularitywith an attractive potential. It is then highly relevant to discuss
the possible resolutions of this singularity before the eventual
decay of the unstable non-BPS branes. One can  test  the 
nature  of  naked singularities  by probing the
supergravity solution with scalars and gravitons. In particular the 
quantum behaviour of gravitons is crucial. In the following we shall apply the
wave-regularity criterion stating that naked singularities with a well-defined
time evolution operator are quantum mechanically sensible\cite{hor,ishi}.

In a first part we will recall the structure of the neutral non-BPS solutions
and show that the naked singularity is a {\it repulson}\cite{kal},
a singularity reflecting perfectly any incoming radiation.
Moreover the wave-regularity criterion is satisfied. Nevertheless
we show that the potential well around the singularity
traps wave-packets preventing the unitarity of the scattering S-matrix.
In a second part we compute the absorption by this potential well
and show that it possesses an infrared singularity obstructing
the decoupling of gravity for non-BPS branes.

\section{Scalar Absorption by Non-BPS Branes}
\subsection{The Scalar Potential}
In the following we shall be interested in the scalar absorption by extended objects\cite{emp}. Let us consider the Klein-Gordon equation for a massless scalar field
in a gravitational background corresponding to $p$-branes
in ten dimensions
\begin{equation}
ds^2_{10}=e^{2A(\rho)}dx_{//}^2 + e^{2B(\rho)}d\rho^2 +e^{2C(\rho)}\rho^2 d\Omega_{8-p}^2
\end{equation}
with Poincar\'e invariance on the brane and rotational invariance in the
transverse direction.
We are considering the scattering of scalars governed by the
equations
\begin{equation}
\frac{1}{\sqrt{g}}\partial_{\mu}(\sqrt{g}g^{\mu\nu}\partial_{\nu}\Phi)=0.
\end{equation}
Let us now study the partial waves 
\begin{equation}
\Phi= e^{-i\omega t}P_l(\Omega)\phi_l(\rho)
\end{equation}
where $P_l$ is an eigenfunction of the spherical Laplacian. 
It is easy to see that the wave equation can be cast in the form of a
Schrodinger equation by putting
\begin{equation}
\phi_l(\rho)=(g^{\rho\rho}\sqrt{g})^{-1/2}F_l(\rho)
\end{equation}
yielding
\begin{equation}
(\frac{d^2}{d\rho^2}-V(\rho))F_l=0
\end{equation}
where the potential reads
\begin{equation}
V=-w^2e^{2B(\rho)-2A(\rho)}+\frac{1}{\sqrt U}\frac{d^2\sqrt U}{d\rho^{2}} + \frac{l(l+7-p)}{\rho^2}e^{2B(\rho)-2C(\rho)}
\end{equation}
and  prime stands for $d/d\rho$. The potential depends on  the form factor  
\begin{equation}
U=g^{\rho\rho}\sqrt{g}.
\end{equation}
The Klein-Gordon equation is covariant under changes of coordinates leading to 
the transformations of the potential and the wave function 
\begin{eqnarray}
\tilde F_l(\rho^*)&=&(\frac{d\rho^*}{d\rho})^{1/2}F_l(\rho)\nonumber \\
\tilde V&=&(\frac{d\rho^*}{d\rho})^{-2}V-\frac{1}{2}\{ \rho,\rho^*\}\nonumber \\
\end{eqnarray}
where the Schwarzian derivative is given by 
\begin{equation}
\{ \rho,\rho^*\}=(\frac{\rho ''}{\rho '})'-\frac{1}{2}(\frac{\rho ''}{\rho '})^2.
\end{equation}
We will mainly use the tortoise coordinates\cite{emp,park}
\begin{equation}
d\rho^*=e^{B(\rho)-A(\rho)}  d\rho
\end{equation}
leading to the s-wave potential
\begin{equation}
V_{tort.}=-\omega^2 + \frac{1}{\sqrt U}\frac{d^2\sqrt U}{d\rho^{*2}}.
\end{equation}
The Schrodinger equation can be written in a factorized form\cite{coh}
\begin{equation}
(\bar Q Q  -\omega^2 )\tilde F_0=0
\end{equation}
where $Q$ and $\bar Q$ are formal adjoints for functions vanishing to second order at the
origin
\begin{equation}
Q=-\frac{d}{d\rho^*}+\frac{1}{2}\frac{d\ln U}{d\rho^*},\ \bar Q=\frac{d}{d\rho^*}+\frac{1}{2}\frac{d\ln U}{d\rho^*}. 
\end{equation}
The Hamiltonien $H=\bar Q Q$ is a symmetric operator $(g,Hf)=(Hg,f)$ where
\begin{equation}
(f,g)=\int d\rho^* f^*(\rho^*) g(\rho^*)+ \int d\rho^* D_{\rho^*}f^*(\rho^*)D_{\rho^*} g(\rho^*)
\end{equation}
for function \footnote{The choice of the norm is crucial for the physical description of the singularities. Following \cite{ishi} we choose a Sobolev norm as it is related
to the energy of the scalar field $\phi$. In particular fields with finite norm have  finite energy.}depending only on $\rho^*$
provided that, at least, the no-flux boundary condition is satisfied
\begin{equation}
f^*(\rho^*)\frac{dg(\rho^*)}{d\rho^*}\vert_{\rho^*=0}=g^*(\rho^*)\frac{df(\rho^*)}{d\rho^*}\vert_{\rho^*=0}.
\end{equation}
This is always satisfied if one restricts the domain $D(H)$ of $H$ to the infinitely differentiable
functions with compact support. 
With this choice the Hamiltonian is symmetric but is not guaranteed to be a self-adjoint operator.

On the other hand
when the flux at the origin is non-zero then unitarity cannot be respected resulting in a
non-zero absorption cross-section. 
Notice too that the Hamiltonian
is a positive operator with two zero modes
\begin{equation}
\psi_1(\rho^*)=\sqrt U,\ \psi_2(\rho^*)= \sqrt U\int ^{\rho^*}\frac{du}{U(u)}.
\end{equation}
The rest of the spectrum is positive preventing the existence of tachyons.

In general it is impossible to find analytic expressions for the wave functions
over the whole space. One has to resort to the matching technique
relating the wave functions in two different patches. The tortoise coordinates
are particularly suited close to the origin and at infinity.

\subsection{Absorption by Non-BPS Branes}
Let us  give the metric in the Einstein frame. It is convenient
to use isotropic coordinates\cite{brax}. The metric reads
\begin{equation}
ds^2_{10}=(\frac{f_-}{f_+})^{\alpha}dx_{//}^2 + f_-^{\beta_-}f_+^{\beta_+}
(r^2d\Omega^2_{8-p}+dr^2)
\end{equation}
where $f_{\pm}=1\pm r_0^{7-p}/r^{7-p}$. The exponents depend on the
tachyon vev $c_1$ as
\begin{eqnarray}
\alpha&=&(7-p) (\frac{-\vert 3-p\vert  c_1+4k}{32} )\nonumber \\
\beta_{\pm}&=&\frac{2}{7-p}\pm (p+1)  (\frac{-\vert 3-p \vert c_1+4k}{32})\nonumber \\
k&=&\sqrt{\frac{2(8-p)}{7-p}-\frac{(p+1)(7-p)}{16}c_1^2}\nonumber \\
\end{eqnarray}
where $\alpha\ge 0$ to guarantee the positivity of the ADM mass. 
It is convenient to use the near singularity coordinate $\rho^{7-p}=r^{7-p}
-r_0^{7-p}$ to describe the space-time for $r\ge r_0$ away from the
naked singularity at $r=r_0$.
In this system of coordinates the metric reads
\begin{equation}
ds^2_{10}=\tilde f_{++}^{-\alpha}dx^2_{//}+\tilde f_+^{2(p-8)/(7-p)}\tilde f_{++}^{\beta_+}(d\rho^2+\rho^2\tilde f_+^2 d\Omega_{8-p}^2 )
\end{equation}
where $\tilde f_+= 1+r_0^{7-p}/\rho^{7-p}$ and $\tilde f_{++}=1+2r_0^{7-p}/\rho^{7-p}$. 
The tortoise coordinates are given by
\begin{equation}
d\rho^*= \tilde f_+^{(p-8)/(7-p)}\tilde f_{++}^{(\alpha+\beta_+)/2}d\rho.
\end{equation}
The metric becomes
\begin{equation}
ds^2_{10}=\tilde f_{++}^{-\alpha}(dx_{//}^2+d\rho ^{*2})+\tilde f_+^{-2/(7-p)}\tilde f_{++}^{\beta_+}\rho^{2}d\Omega_{8-p}^2
\end{equation}
leading to
\begin{equation}
U=\tilde f_+^{(p-8)/(7-p)}\tilde f_{++}^{(\alpha  +
\beta_+)/2 +1 }\rho^{8-p}.
\end{equation}
Close to the singularity it is convenient to study the behaviour of $U$
and $\rho^*$ as a function of $\rho$
\begin{eqnarray}
U(\rho)&\sim& S\rho^{-a}\nonumber \\
\frac{d\rho^*}{d\rho}&\sim& T \rho^{-1-a}\nonumber\\
\end{eqnarray}
where the constants are defined by
\begin{equation}
a= p-9+(7-p)\frac{\alpha+\beta_+}{2}
\end{equation}
and $S= 2^{-p\alpha/2+(8-p)\beta_+/2}r_0^{p-8-(7-p)(p\alpha-(8-p)\beta_+)/2},\ 
T=2^{(\alpha+\beta_+)/2}r_0^{p-8+(7-p)(\alpha+\beta_+)/2}$.
Notice that $a<0$ implying that $\rho^*$ vanishes for vanishing $\rho$. This implies that 
the naked singularity at the origin is time-like.
The resulting Schrodinger equation is of the Bessel type of zeroth order with generalized eigenvalues 
\begin{equation}
\tilde \psi_{\omega}^1(\rho^*)=\sqrt\rho^*J_0(\vert \omega\vert  \rho^*),\ \tilde \psi_{\omega}^2
=\sqrt \rho^* N_0(\vert \omega\vert  \rho^*).
\end{equation}

Close to the singularity the constant term in the potential is negligible implying that all the solutions
behave like the two zero modes
\begin{equation}
\psi_1(\rho^*)=\sqrt \rho^*,\ \psi_2(\rho^*)= \sqrt \rho^*\ln \rho^*.
\end{equation}
It is easy to see that  
no flux reaches the singularity and the s-wave
absorption cross section by the singularity vanishes exactly.
The same result holds for the higher partial waves.

\subsection{Naked Singularities}

The previous result sheds some light on the nature of the naked singularity at the origin.
Classically the naked singularity is attractive as can be seen from the form of the potential
\begin{equation}
V_p(\rho^*)=-\omega^2 -\frac{1}{4\rho^{*2}}
\end{equation}
close to the origin. 
Quantum mechanically the potential is so steep that
the scalar probes are totally reflected. The singularity is then a {\it repulson}\cite{kal}.
One can even go deeper in the analysis of the singularity by describing  the time evolution of scalar probes
in the vicinity of the singularity\cite{hor,ishi}. To do that let us write the massless Klein-Gordon equation 
in the form
\begin{equation}
\frac{\partial^2 \phi}{\partial t^2}=- M \phi
\end{equation}
where $M$ is a second order partial differential operator depending only on the spatial derivatives.
After a change of variable, $M $ reduces to the Hamiltonian $H$. 
The Klein-Gordon equation defines a unique time evolution provided it can be written in the Schrodinger form
\begin{equation}
\frac{\partial \phi}{\partial t}=i M^{1/2} \phi
\end{equation}
for a unique self-adjoint operator $M^{1/2}$. 
This is equivalent to finding a unique self-adjoint extension to the symmetric operator $H$, i.e.
the Hamiltonian $H$ is essentially self-adjoint.

There is a useful criterion
of essential self-adjointness.
Let us consider the eigenvalue problem
\begin{equation}
\nabla^{\mu}\nabla_{\mu}\phi=\pm i\phi
\label{index}
\end{equation}
in the non-BPS background. It reduces to a Schrodinger problem
in a complex potential
\begin{equation}
V=V_p\pm ig_{\rho\rho}.
\end{equation}
Denote by $n_{\pm}$ the number of normalizable solutions  to (\ref{index}).
As the operator $H$ is real one has $n_+= n_-$ implying that there always exists 
self-adjoint extensions. Now the operator is essentially self-adjoint provided
$n_{\pm}=0$, i.e. the solutions are not normalizable.
This is achieved provided that one of the solutions for each sign in (\ref{index}) is
non-normalizable\cite{hor}. 

In our case notice that in the vicinity of the singularity the extra complex term to $V_p$ is  negligible
implying that the solutions are expressed in terms of the two zero modes. 
The issue of the quantum mechanical behaviour  of the singularity is
now dependent on the norm of these  eigenfunctions. 
We find that the norm of $\psi_2(\rho^*)$ is  divergent. Therefore the Hamiltonian $H$ is essentially self-adjoint.

Physically the naked singularity at the origin is a repulson leading  a well-defined
time evolution of scalar probes. This does not prevent the S-matrix from  being non-unitary.
Indeed let us consider the time evolution of a wave packet initially at
infinity in the $\rho^*$ direction. Noting that the eigenfunctions at infinity
behave like free waves $e^{\pm i\omega \rho^*}$, we can expand this initial
wave packet on the eigenvectors of $H$  
\begin{equation}
\tilde F_0(\rho^*,0) =\int d\omega  a(\omega)( \psi_{\omega}^1(\rho^*)+i \psi_{\omega}^2(\rho^*)).
\end{equation}
The two eigenvectors behave like $ \psi_{\omega}^1(\rho^*)=i
\sqrt{\frac{2}{\pi\vert\omega\vert}}\cos (\vert \omega\vert  \rho^*-\pi/4)$
and $ \psi_{\omega}^2(\rho^*)= i
\sqrt{\frac{2}{\pi\vert\omega\vert}} \sin (\vert \omega\vert  \rho^*-\pi/4)$ at infinity.
At any given time the wave packet becomes 
\begin{equation}
\tilde F_0(\rho^*,t) =\int d\omega e^{i\omega t} a(\omega)( \psi_{\omega}^1(\rho^*)+i \psi_{\omega}^2(\rho^*))
\end{equation}
evolving towards the singularity.
Let us specialize to the neighbourhood of the singularity
where  all the generalized eigenfunctions are $\omega$-independent 
\begin{equation}
\tilde F_0(\rho^*,t)=(\int d\omega 
a(\omega)e^{iwt})\ (\psi_1 (\rho^*)+i \psi_2(\rho^*)).
\end{equation}
For a very large time corresponding to the very large initial value of $\rho^*$
at the centre of  the wave packet 
we find that the wave packet reaches the neighbourhood of the singularity.
Moreover the wave packet becomes 
non-normalizable due to the $\psi_2(\rho^*)$ component. Therefore the infinite time limit of the time
evolution operator is not unitary.  This is due to the fact that the singularity is
surrounded by an absorbing infinite well.

In a sense this provides a quantum mechanical resolution of the naked singularity. Indeed
the classically  attractive singularity is replaced quantum-mechanically
by a finite absorbing region surrounding a repulson.

\section{Graviton Absorption}
\subsection{Gravitons vs. Scalars}
Let us analyse the absorption of gravitons polarized along the brane.
This corresponds to perturbations of the metric of the brane  by the
incoming gravitons.
In the Einstein frame the graviton equation reduces to the Laplace  equation
\begin{equation}
\Delta h_{ij}=0.
\end{equation}
Consider  waves of the form
\begin{equation}
h_{ij}=\epsilon_{ij} h e^{ik.x}
\end{equation}
where $\epsilon$ is the polarization tensor and $k$ is in the time direction.
The polarization tensor must be traceless $\eta^{ij}\epsilon_{ij}=0$
and transverse to $k$ implying that $\epsilon_{0i}=0$. Denoting
by $\tilde\epsilon$ the spatial part of the polarization tensor we find
a basis of these tensors given by off-diagonal symmetric matrices
with $\tilde \epsilon_{ab}=\tilde \epsilon_{ba}=1$ and zero otherwise
along with diagonal matrices such that $ \tilde \epsilon_{aa}=1,\
\tilde \epsilon_{bb}=-1,\ a<b$. The latter are diagonal
polarizations while the former are transverse polarizations.
In the diagonal case put
\begin{equation}
h=g_{\rho\rho}^{1/2}\phi.
\end{equation}
Then the scalar field $\phi$ satisfies the free wave equation
\begin{equation}
\nabla_{\mu}\nabla^{\mu} \phi=0.
\end{equation}
In the transverse case the function $h$ satisfies the free scalar equation too.

We will calculate the absorption of gravitons by the non-BPS branes.
It is convenient to define the conserved flux\cite{ali} 
\begin{equation}
{\cal F}_{grav}=\frac{1}{2i}\int \sqrt{g}g^{\mu\nu}g^{\alpha\beta}g^{\rho\rho}
(h^*_{\mu\alpha}\partial_{\rho}h_{\nu\beta}-\partial_{\rho}h^*_{\mu\alpha}
h_{\nu\beta}).
\end{equation}
The flux becomes 
\begin{equation}
{\cal F}_{grav}=\frac{1}{2i}\int \sqrt{g}g^{ab} g^{cd}\tilde\epsilon_{ac}
\tilde\epsilon_{bd}g^{\rho\rho}
(h^*\partial_{\rho}h-\partial_{\rho}h^*
h).
\end{equation}
which is directly related to the scalar flux.
In the following we shall compute the scalar absorption.

\subsection{Non-BPS Graviton Absorption}

The scattering potential requires the evaluation of $g^{\rho\rho}\sqrt{g}=
\rho^{8-p}\tilde f_{++}$. This leads to the form factor
\begin{equation}
U=\rho^{8-p}+2r_0^{7-p}\rho.
\end{equation}
The potential follows
after rescaling $z=\rho \omega$
\begin{equation}
V_p=-(1+\frac{2\mu}{z^{7-p}})^{\alpha+\beta_+}(1+\frac{\mu}{z^{7-p}})^{2(p-8)/(7-p)}+\frac{(8-p)(6-p)}{4z^2}- \mu^2\frac{ (7-p)^2}{z^2(z^{7-p}+2\mu)^2}
\end{equation}
where $\mu=(\omega r_0)^{7-p}$. For higher partial waves there is an extra repulsive contributions 
\begin{equation}
\frac{l(l+7-p)}{\rho^2 \tilde f_{+}^{2}}.
\end{equation}
The potential has only a weak dependence on $c_1$.

Notice that the shape of the potential is different for $p=6$ and $p<6$.
Let us first focus on the $p=6$ case.
\begin{itemize}
\item {\bf p=6}
\end{itemize}
We can distinguish three different regions.
In the core  for $z<<\mu $ the potential reads
\begin{equation}
V_6=-1-\frac{1}{4z^2}.
\end{equation}
This is the universal behaviour of the previous section which also appears
for the naked singularities of cosmic string solutions\cite{coh}.
Moreover the extra contribution from the
higher partial waves is negligible compared to the $1/z^2$ behaviour.

The absorption cross section is obtained by considering incoming waves
reaching the outer boundary of the core region $\rho\approx r_0$ and
comparing this  incoming flux to the one  at infinity. 
To do so we need to study the potential 
outside the core  
\begin{equation}
V_6= -(1+\frac{4k-3c_1}{2}\frac{\mu}{z}) -\frac{\mu^2}{z^4}.
\end{equation}
In the  inner region $\mu << z << \mu^{1/3}$ for $\mu  <<1 $
the potential becomes 
\begin{equation}
V_6=-1-\frac{\mu^2}{z^4}.
\end{equation}
Finally in the outer region  $z>>\mu^{1/3}$ the potential reads
\begin{equation}
V_6=-1-\frac{4k-3c_1}{2}\frac{\mu}{z}.
\end{equation}
In the outer region the solutions are known to be expressible in terms of  confluent hypergeometric functions\cite{ali} 
\begin{equation}
F_0={\cal A}_6ze^{iz}\ _1F_1(1-i\frac{4k-3c_1}{4}\mu,2;-2iz).
\end{equation}
The amplitude ${\cal{A}}_6$ is obtained by matching with the inner region where
the solution can be expressed in terms of Mathieu functions. It will be more
transparent to use a duality transformation reducing the solutions to
Hankel functions. 
Define
\begin{equation}
\tilde z=\frac{\mu}{z}
\end{equation}
and 
\begin{equation}
F_0(z)=\tilde z^{-1/2}\tilde F_0(\tilde z).
\end{equation}
The equation in the dual variables is now
\begin{equation}
\frac{d^2\tilde F_0}{d\tilde z^2}+\tilde z\frac{d\tilde F_0}{d\tilde z}+
(1-\frac{1}{4\tilde z^2}+\frac{\mu^{2}}{\tilde z^{4}})\tilde F_0=0
\end{equation}
which is a Bessel equation for $z <<1/2$. There is an overlapping interval  
for the  solution of the dual equation
and the solution in the outer region.
More precisely the solution is expressed in terms of the  Hankel function
$H^{(1)}_{1/2}$ and reads
\begin{equation}
F_0=i\sqrt{\frac{2}{\pi}}\frac{z}{\mu}
e^{i\mu/z}
\end{equation}
for $\mu<<z<<1/2$. In this interval the arguments
of the Hypergeometric and Hankel functions are small. 
The inner function behaves like $i\sqrt{\frac{2}{\pi}}\frac{z}{\mu}$ while the 
hypergeometric function matches this behaviour 
implying that 
\begin{equation}
{\cal A}_6= i\sqrt{\frac{2}{\pi}}\frac{1}{\mu}.
\end{equation}
This specifies entirely the scalar wave function outside the core
of the brane. 

We can now evaluate the absorption cross section. To do so one needs to evaluate the ingoing fluxes
at infinity and at the outer boundary of the core, i.e. $r=2r_0$. 
The scattering cross section is given by
\begin{equation}
\sigma_6=\frac{4\pi^2}{\Omega_2 w^2}\frac{{\cal F}^{in}_{2r_0}}{{\cal F}_{\infty}^{in}}
\end{equation}
where $\Omega_d=2\pi^{(d+1)/2}/\Gamma({(d+1)/2})$.

Notice  that the incoming wave  for small $\mu$ behaves like
${\cal A}_6 e^{iz}$ at infinity. We obtain the flux  per unit volume $ {{\cal F}_{\infty}^{in}}=
4\pi \omega \vert {\cal A}_6\vert ^2 $. 
Similarly the flux at $2r_0$ is given by ${\cal F}_{2r_0}^{in}=8/r_0$.
Now the corresponding
absorption cross section is given by
\begin{equation}
\sigma_6=\frac{\pi r_0}{\omega}.
\end{equation}
The cross section diverges for small $\omega$.

As already seen in \cite{brax} there is no decoupling limit of the brane modes
from the bulk modes. In particular this can be shown  by considering a non-BPS brane
whose size scales with the string length and computing the absorption cross section in string units
in the limit $l_s \to 0$. Dimensionally the radius $r_0$ is given by
$r_0\propto  l_s$. In the small string length limit
the ratio
\begin{equation}
\frac{\sigma_6}{l_s^2}\propto \frac{\pi}{l_s \omega}
\end{equation}
measures the absorption in string units. 
As $l_s\omega << 1$ we find that the scalars and therefore the gravitons
do not decouple from the brane.

\begin{itemize}
\item  ${\bf p \ne  6 }$
\end{itemize}

We can distinguish three different regions.
In the core  for $z<<\mu^{1/(7-p)}$ the potential reads
\begin{equation}
V_p=-1-\frac{1}{4z^2}.
\end{equation}
This is the near singularity behaviour that has already been 
observed in the previous section. Moreover the extra contribution from the
higher partial waves is negligible compared to the $1/z^2$ behaviour.

Let us now study the behaviour outside the core. 
In the inner region  $\mu^{1/(7-p)}<<z<< \mu^{1/(9-p)}$
the potential reads
\begin{equation}
V_p=-1+\frac{(8-p)(6-p)}{4z^2}- \mu^2\frac{ (7-p)^2}{z^{2(8-p)}}.
\end{equation}
Finally in the outer region  for $z>>\mu^{1/(9-p)}$
the potential decreases like
\begin{equation}
V_p=-1+ \frac{(8-p)(6-p)}{4z^2}.
\end{equation}
On the last expression one can see that the potential decreases at infinity,
as it increases around the origin there is a maximum for 
$z_*=O(\mu^{1/7-p})$ of height $V_p(z_*)=O(\mu^{-2/(7-p)})$.
This is highly similar to the extremal case\cite {ali} where there is also a maximum for
the potential when $p<6$. 
The presence of a maximum in the potential should not be mistaken with
a barrier preventing the penetration of the incoming waves. Indeed if one uses a more appropriate spherical representation of the Schrodinger equation one gets 
 by putting $F_0=z^{(8-p)/2}f_0$ 
\begin{equation}
-\frac{1}{z^{8-p}}\frac{d}{dz}(z^{8-p}\frac{df_0}{dz})+ \tilde V_pf_0=0
\end{equation}
where
\begin{equation}
\tilde V_p=V_p-\frac{(8-p)(6-p)}{4z^2}.
\end{equation}
The additional contribution cancels exactly the repulsive part of the potential
leading to an attractive potential altogether.

Let us now concentrate on the absorption .
To do so it is convenient to solve first the Schrodinger in dual variables
\begin{equation}
\tilde z=\frac{\mu}{z^{7-p}}
\end{equation}
and 
\begin{equation}
F_0(z)=\tilde z^{-1/2(7-p)}\tilde F_0(\tilde z).
\end{equation}
The equation in the dual variables is now
\begin{equation}
\frac{d^2\tilde F_0}{d\tilde z^2}+\tilde z\frac{d\tilde F_0}{d\tilde z}+
(1-\frac{1}{4\tilde z^2}+\frac{\mu^{2/(7-p)}}{(7-p)^2 \tilde z^{2+2/(7-p)}})\tilde F_0=0
\end{equation}
which is a Bessel equation for $z <<(7-p)/2$. There is an overlapping interval  for the validity of the solution of the dual equation
and the solution in the outer region.
More precisely the solution reads now
\begin{equation}
F_0=i\sqrt{\frac{2}{\pi}}\frac{z^{(8-p)/2}}{\mu^{(8-p)/2(7-p)}}
e^{i\mu/z^{7-p}}
\end{equation}
for $\mu^{1/(7-p)}<<z<<(7-p)/2$.
In the outer region the solution is a Bessel function of order $(7-p)/2$
\begin{equation}
F_0={\cal A}_pz^{1/2}J_{(7-p)/2}(z).
\end{equation}
Now for small arguments $F_0\sim {\cal A}_p \frac{z^{(8-p)/2}}
{2^{(7-p)/2}\Gamma((9-p)/2)}$
implying that
\begin{equation}
{\cal A}_p= i\sqrt{\frac{2}{\pi}}{2^{(7-p)/2}}{\Gamma(\frac{9-p}{2})}\mu^{(p-8)/2(7-p)}.
\end{equation}
We can now deduce the flux at infinity
\begin{equation}
{\cal F}_{\infty}^{in}=\Omega_{8-p}
\frac{\omega }{2\pi} \vert {\cal A}_p\vert^2.
\end{equation}
Similarly the flux close the  core is
\begin{equation}
{\cal F}_{r_*}^{in}=
\Omega_{8-p}\frac{2(7-p)\omega}{\pi}\mu^{-1/(7-p)}
\end{equation}
where $r_*=2^{1/(7-p)}r_0$.
This leads to the cross section
\begin{equation}
\sigma_p= \frac{(2\pi)^{8-p}}{\omega^{8-p}\Omega_{8-p}}\frac{{\cal F}_{r_*}^{in}}{{\cal F}_{\infty}^{in}}
\end{equation}
which reads 
\begin{equation}
\sigma_p= \frac{(7-p)\Omega_{8-p}r_0^{7-p}}{\omega}.
\end{equation}
The cross-section is divergent for soft probes.

Let us now consider a non-BPS brane with $r_0\propto l_s$ and study the decoupling
of gravity in the small string length regime.
The absorption cross section behaves like 
\begin{equation}
\frac{\sigma_p}{l_s^{8-p}}\propto \frac{(7-p)\Omega_{8-p}}{ l_s\omega}
\end{equation}
implying that the gravitons do not decouple from the inner region.

Another interesting point is the $c_1$ independence of $\sigma_p$.
 This entails that the probes are insensitive to the inner state
of the brane, i.e. whether the tachyon possesses a vev. 

Notice that we expect that higher curvature terms in the supergravity Lagrangian will
modify the non-BPS solutions. Nevertheless the previous absorption cross section has been
obtained from incoming waves penetrating the high curvature region where absorption takes place.
Outside this region the non-BPS solutions should remain valid and therefore the
absorption cross-section should keep the same features, i.e. the absence of decoupling of gravity.

\section{Conclusions}
We have studied the absorption of scalars and gravitons by non-BPS branes.
One of the main features of non-BPS branes is the existence of a naked singularity.
From the classical point of view this is an attractive singularity 
which needs to be resolved. Quantum mechanically when probed by gravitons and scalars
the naked singularity appears to be a repulson reflecting incoming waves. Nevertheless the
potential well around the repulson is so steep that wave-packets are trapped.
This entails that the scattering S-matrix is not unitary. The absorption cross-section
possesses an infra-red singularity which signals the non-decoupling of gravity.

In summary we have found that non-BPS branes are of a peculiar nature.
Indeed the graviton probes   do not reach the naked singularity 
while they remain  strongly coupled to the 
neighbourhood of this  singularity.
Perhaps this is the supergravity way of observing that at short distances
the predominant features
of non-BPS branes spring from open string excitations.
This suggests a possible resolution of the naked singularity by a non-supersymmetric {\it enhan\c con} mechanism\cite{john,berg} whereby the $\rho \le r_0$ region is smoothed
out while the $\rho \ge r_0$ remains unchanged.


\begin{thebibliography}{999}


\bibitem{senreview} A. Sen, ``Non-BPS states and Branes in String
  Theory'', APCTP winter school lectures, hep-th/9904207.

\bibitem{sentach} A. Sen, JHEP 12, 021 (1998).


\bibitem{numerics} A. Sen, B. Zwiebach, { JHEP} {\bf 0003} (1999) 002; N.
  Berkovits, A. Sen and B. Zwiebach, Nucl. Phys. {\bf B587} (2000) 022.

\bibitem{Harvey-Horava-Kraus} J. A. Harvey, P. Horava, P. Kraus, 
 { JHEP} 0003 (20002) 021.


\bibitem{das} K. Dasgupta, S. Mukhi and G. Rajesh, JHEP {\bf 0006}, 022 (2000).

\bibitem{har} J. A. Harvey, P. Kraus, F. Larsen and E. J. Martinec, JHEP {\bf 0007}, 042 (2000).

\bibitem{man} G. Mandal and S. R. Wadia, ``Matrix Model, Noncommutative Gauge Theory and the Tachyon Potential'', hep-th/0011094.

\bibitem{intri} K. Intriligator, M. Kleban and J. Kumar, ``Comments on Unstable Branes'', hep-th/101010.

\bibitem{ler} M. Bertolini, P. Di Vecchia, M. Frau, A. Lerda, R, Marotta and R. Russo, Nucl. Phys. {\bf B590} (2000), 471.

\bibitem{brax} Ph. Brax, G. Mandal and Y. Oz, ``Supergravity Description of Non-BPS Branes'', hep-th/0005242.

\bibitem{hor} G. T Horowitz and D. Marolf, Phys. Rev. {\bf D 52} (1995) 5670

\bibitem{ishi} A. Ishibashi and A. Hosoya, Phys. Rev. {\bf D 60} (1999) 104028


\bibitem{kal} R.. Kallosh and A. Linde, Phys. Rev. {\bf D 52} (1995) 7137.


\bibitem{emp} R. Emparan,  Nucl. Phys. {\bf B516} (1998)  297.


\bibitem{park} D. K. Park and H. J. W. Muller-Kirsten, Phys. Lett. {\bf B492} (2000) 135.


\bibitem{coh} A. G. Cohen and D. K. Kaplan, Phys. Lett. {B470} (1999) 52.


\bibitem{ali} M. Alishahiha, H. Ita and Y. Oz, JHEP 0006 (2000) 002.


\bibitem{john} C. V. Johnson, A. W. Peet and J. Polchinski, Phys. Rev. {\bf D 61} (2000) 0860001.


\bibitem{berg} P. Berglund, T. Hubsch and D. Minic, ``Probing Naked Singularities in Non-Supersymmetric String Vacua'', hep-th/0012042.


\end{thebibliography}
\end{document}